\documentclass[preprint,showkeys,nofootinbib]{revtex4}
\usepackage{amssymb,upref,graphicx,lastpage}

\usepackage{amsfonts}
\usepackage{amsmath}
\usepackage{amssymb}
\usepackage{graphicx}
\usepackage{subfigure}

\begin{document}

\title{A density-functional theory investigation of cluster formation in an effective-potential model of dendrimers}

\author{D. Pini}
\affiliation{Universit\`a degli Studi di Milano, Dipartimento di Fisica, Via Celoria 16, 20133 Milano, Italy}
\email{davide.pini@fisica.unimi.it}


\begin{abstract}
We consider a system of particles interacting via a purely repulsive, soft-core potential recently introduced to model
effective pair interactions between dendrimers, which is expected to lead to the formation of crystals with 
multiple occupancy of the lattice sites. The phase diagram is investigated 
by density-functional theory (DFT) without making any {\em a priori} assumption on the functional form of the density 
profile
or on the type of crystal lattice. As the average density $\rho$ is increased, the system displays first a transition 
from a fluid to a bcc phase, and subsequently to hcp and fcc phases. In the inhomogeneous region, the behavior 
is that found in previous investigations of this class of cluster-forming potentials. Specifically, the particles 
arrange into clusters strongly localized at the lattice sites, and the lattice constant depends very weakly on $\rho$, 
leading to an occupancy number of the sites which is a nearly linear function of $\rho$. 
These results are compared to those predicted by the more widespread approach, in which the DFT minimization 
is carried out by representing the density profile by a given functional form depending on few variational parameters. 
We find that for the model potential studied here, the latter approach recovers most of the predictions 
of the unconstrained minimization.

\end{abstract}

\maketitle

\section{Introduction}
\label{sec:introduction}

Bounded repulsive potentials arise in the description of effective, coarse-grained 
interactions between polymers. Because of their noncompact nature, mutual overlap 
between polymers does not require the actual superposition between 
the constituent monomers and is thus allowed, while still implying a steric repulsion 
due to the reduction of their available configurations. The prototypical example
of these effective {\em soft-core} interactions is the repulsive Gaussian two-body 
potential, whose thorough study~\cite{stillinger} has shown how its equation of state, correlations,
and phase diagram differ markedly from those of hard-core fluids.  

In recent years, much attention has been given to another class of soft-core repulsive 
interactions which occur, for instance, in the modelization of ring polymers~\cite{narros} 
or dendrimers~\cite{mladekrev2,lenz}.  
These interactions differ from the simple Gaussian potential mainly for being 
softer at small interparticle separation $r$, possibly displaying even  
a minimum at $r=0$. As a consequence, their Fourier transform is not  
positive everywhere, and has a negative minimum at non-vanishing wavevector $k_{\rm m}\neq 0$ --- a property sometimes 
referred to as characteristic of the $Q^{\pm}$ class of repulsive interactions~\cite{likosq}, which causes the tendency
for the system to self-assemble into regular structures consisting of clusters of particles with
a period $\sim 2\pi/k_{\rm m}$, basically independent of density~\cite{likoschem}. This scenario is similar to that 
giving rise to microphases in hard-core fluids with competing attractive and repulsive interactions~\cite{compe,archer}, 
although the physical origin of clustering is different in the two cases. In fluids with competing interactions, 
clusters occur because on the one hand the attraction favors particle aggregation, while on the other hand
the repulsion prevents the aggregates from growing beyond a certain size. In $Q^{\pm}$ repulsive potentials,
instead, clustering is due to the fact that particles are allowed to superimpose to each other at the price 
of a very small mutual repulsion.      

The formation of cluster phases in $Q^{\pm}$ potentials has been investigated in detail for the 
generalized exponential model (GEM) with interaction of the form~\cite{likoschem,likosrev}:  
\begin{equation}
v_{\rm GEM}(r)=\epsilon\exp[-(r/R)^{4}] \, .
\label{gem}
\end{equation}
In this case one finds that, when the density is increased at sufficiently high temperature, the homogeneous
fluid is replaced first by a cluster bcc phase, then by a cluster fcc phase at higher density.
Below a triple-point temperature $T_{\rm t}$, the bcc phase
disappears, and the only transition left is that between the fluid and the fcc cluster phase~\cite{likosrev,mladekrev} 
until, at even lower temperatures, the fcc phase splits into a sequence of fcc phases connected to each other by 
isostructural transitions, in which the lattice constant jumps discontinuously~\cite{likosj,zhang}.  

\begin{figure}
\includegraphics[width=9cm]{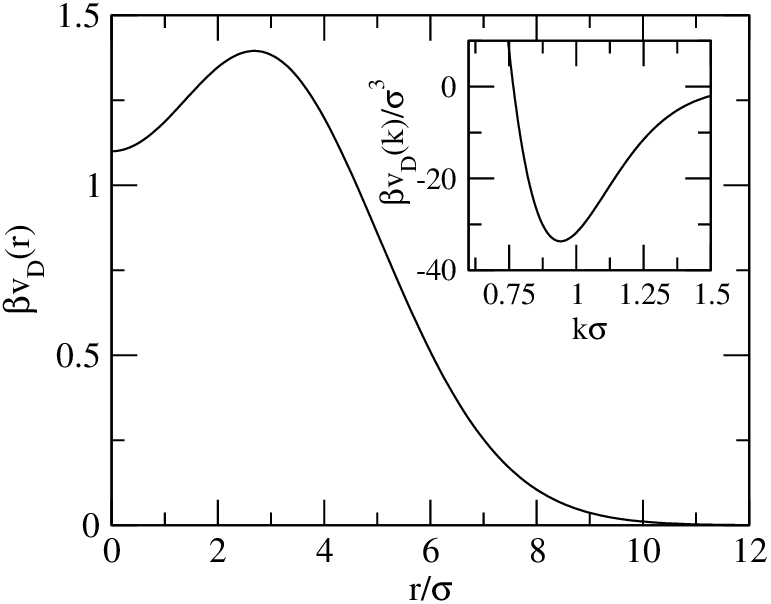}
\caption{A plot of the potential $v_{\rm D}(r)$ of Eq.~(\ref{dendr}) for the interaction between 
two isolated dendrimers. The inset 
shows the Fourier transform of the potential $\beta\tilde{v}_{\rm D}(k)$ in the neighborhood of the wave vector 
$k_{\rm m}$ at which it attains its minimum.}
\label{fig:dendr}
\end{figure}

Another family of potentials of this class has been proposed as a modelization of the effective interactions 
between amphiphilic dendrimers with a solvophobic core and a solvophilic shell~\cite{mladekrev2,lenz}. In particular, 
for a specific choice of the parameters which determine 
the potentials between adjacent and non-adjacent monomers of the dendrimer, the interaction 
between two isolated dendrimers can be accurately represented by a fit to a sum of two Gaussians of opposite
sign~\cite{lenz}:
\begin{equation}
\beta v_{\rm D}(r)=\epsilon_{1}\exp[-(r/R_{1})^{2}]-\epsilon_{2}\exp[-(r/R_{2})^{2}] \, ,
\label{dendr}
\end{equation}
where $\beta=1/(k_{\rm B}T)$, $k_{\rm B}$ being the Boltzmann constant, and $\epsilon_{1}=23.6$, 
$\epsilon_{2}=22.5$, $R_{1}=3.75\sigma$, $R_{2}=3.56\sigma$, $\sigma$ being the diameter 
of the core monomers. A plot of this potential is displayed in Figure~\ref{fig:dendr}. Its
Fourier transform $\beta\tilde{v}_{\rm D}(k)$ in the neighborhood of the minimum is shown in the inset. 
The main qualitative difference between $v_{\rm GEM}(r)$ and $v_{\rm D}(r)$
is that in the latter, $r=0$ corresponds to a minimum rather than a maximum, 
and the maximum is assumed for $r\neq 0$. This is due to the fact that at small separation 
between the dendrimers, the overlap between the solvophobic cores contrasts that between the solvophilic 
shells, thereby leading to a decrease in the repulsion with respect to that felt at intermediate separation 
when the cores do not overlap, but the shells still do.    

Even though the formation of cluster crystals in dendrimers has already been addressed by computer 
simulation~\cite{lenz2}, to our knowledge the phase diagram of a system of effective particles interacting by 
the pair potential~(\ref{dendr}) has not been considered yet. In the present work, we address this problem  
by density-functional theory (DFT). We have used a simple, mean-field like free-energy functional, which 
has already been adopted many times to describe cluster formation in both the soft-core 
repulsion~\cite{likoschem,likosrev,mladekj} 
and the competing interaction scenario~\cite{archer}. Here, we have given special attention to the procedure 
followed to minimize the functional. In fact, instead of describing the density profile 
by a given analytical expression depending on a small number 
of free parameters, we have not made any {\em a priori} hypothesis as to its form 
save for assuming that it is periodic, and the Bravais lattice vectors which determine the periodicity 
have themselves been obtained as a result of the minimization. Such a method is computationally more demanding
than that more commonly adopted, but its unbiased nature is an asset, since it does not require 
any guess on  
the structures into which the system is most likely to assemble. 

The plan of the paper is as follows: in Sec.~\ref{sec:theory} we briefly introduce the free-energy 
functional which we have used, and describe the method by which it has been minimized; 
in Sec.~\ref{sec:results} we apply this method to the study 
of the phase diagram of a system of dendrimers, described as effective particles interacting via 
the potential~(\ref{dendr}); in Sec.~\ref{sec:gaussian}, we try to get further insight into the results 
discussed in Sec.~\ref{sec:results}, by assessing up to which extent they can be recovered by adopting an analytical
representation of the density profile; finally, in Sec.~\ref{sec:concl} we summarize our results and draw 
our conclusions.

\section{Theory}
\label{sec:theory}

In order to study the occurrence and structure of the inhomogeneous phases of a system of particles 
interacting via the pair potential~(\ref{dendr}), we have used DFT. According to DFT, at a given 
temperature $T$, chemical potential $\mu$, and volume $V$, the density profile is obtained
by minimizing the grand potential functional $\Omega[\rho({\bf r})]$, whose value then yields 
the grand potential $-PV$, where $P$ is the pressure. 
Henceforth, the density profile will be denoted by $\rho({\bf r})$, while $\rho$ with no point-dependence indicated
will refer to the average density $\rho\equiv\int\!\!d^{3}{\bf r}\rho({\bf r})/V$. 
For a system of particles interacting via a pair potential $v({\bf r})$, the exact 
grand potential functional is of course unknown, and many approximate formulations of DFT have been 
suggested. Here we have used the following simple form for $\Omega[\rho({\bf r})]$:
\begin{equation} 
\beta\Omega=\int\!\!d^{3}{\bf r} \, 
\rho({\bf r})\!
\left\{\ln[\rho({\bf r})\Lambda^{3}]-1-\beta\mu\right\} \, + \, 
\frac{1}{2}\int\!\!d^{3}{\bf r}\!\!\int\!\!d^{3}
{\bf r}^{\prime}
\rho({\bf r})\rho({\bf r}^{\prime})
\beta v({\bf r}-{\bf r}^{\prime}) \, ,
\label{dft}
\end{equation}
where $\Lambda=h/\sqrt{2\pi m k_{\rm B}T}$ is the thermal wavelength of a particle of mass $m$. In the rest of the paper
$\Lambda$ will be dropped, since its actual value does not affect any of the results presented here. 
The first term in the rhs of Eq.~(\ref{dft}) is the exact ideal-gas contribution, while the interaction
is approximately taken into account in the second, ``excess'' term. 
This expression can be considered as 
a generalization to inhomogeneous systems of the van der Waals mean-field approach, which is 
in fact obtained from Eq.~(\ref{dft}) in the special case of a uniform density profile 
$\rho({\bf r})\equiv\rho$.  
A similar approximation for $\Omega[\rho({\bf r})]$ could be adopted also in a system of non-overlapping 
particles whose pair interaction contains a singular repulsive part, but this would require 
as a preliminary step to treat 
separately the singular contribution. Here, instead, $v({\bf r})$ coincides with
the total soft-core interaction $v_{\rm D}(r)$ of Eq.~(\ref{dendr}).  

For a homogeneous state, the stability condition with respect to a small perturbation of the density 
$\delta\rho({\bf r})$   
\begin{equation}
\int\!\!d^{3}{\bf r}\!\!\int\!\!d^{3} {\bf r}^{\prime} \!\left.\frac{\delta^{2}(\beta\Omega)}
{\delta\rho({\bf r})\delta\rho({\bf r}')}\right|_{\rho}\delta\rho({\bf r})\delta\rho({\bf r}')
> 0
\label{stabi}
\end{equation}
is equivalent to the requirement $S(k)>0$ for every wave vector $k$, where 
$S({\bf k})$ is the structure factor. Equation~(\ref{dft}) gives the random-phase approximation (RPA)
for $S(k)$:
\begin{equation}
S(k)=[1+\rho\beta\tilde{v}(k)]^{-1} \, ,   
\label{struct}
\end{equation}
where $\tilde{v}(k)$ is the Fourier transform of $v(r)$. If $v(r)$ belongs to the $Q^{\pm}$ class, 
the above condition on $S(k)$ is violated for $\rho>1/[\, \beta|\tilde{v}(k_{\rm m})|\, ]$, where $k_{\rm m}$ 
is the wave vector at which $\tilde{v}(k)$ assumes its negative minimum. The boundary between the
region where the homogeneous state is stable and that where it is not corresponds to the condition
of marginal stability $\rho=1/[\, \beta|\tilde{v}(k_{\rm m})|\, ]$, which identifies a locus in the $T$-$\rho$ 
plane generally referred to as the $\lambda$-{\em line}. Along the $\lambda$-line, the RPA $S(k)$ diverges 
at $k_{\rm m}$, thus signalling the tendency of the system to form density modulations with a characteristic
length $\sim 2\pi/k_{\rm m}$. This can be compared with the behavior on the spinodal line 
predicted by the RPA for systems which undergo a liquid-vapor transition: in the latter case, 
the divergence of $S(k)$ occurs for $k=0$, and the homogeneous phase is unstable with respect to separation 
into two bulk phases.         

In the present case in which the potential is athermal, the $\lambda$-line is actually 
a line at constant $\rho=\rho_{\lambda}$. For the potential~(\ref{dendr}) one has 
$k_{\rm m}\sigma=0.941$, 
$\beta\tilde{v}_{\rm D}(k_{\rm m})/\sigma^{3}=-33.686$,
so that $\rho_{\lambda}\sigma^{3}=0.030$. However, it should be recalled that, just
like the spinodal line for the liquid-vapor transition, the $\lambda$-line does not generally coincide with
the boundary between the homogeneous and the inhomogeneous fluid. The latter must be obtained by comparing
the free energies of the two phases, and may (and usually will) lie in a region where the homogeneous fluid
is still mechanically stable. 

In order to obtain further insight in the phase diagram and the structure of the inhomogeneous phases, 
it is necessary to turn to the minimization of the functional~(\ref{dft}).   
As stated in Sec.~\ref{sec:introduction}, we have assumed from the outset that the density profile $\rho({\bf r})$ 
is periodic, i.e.
\begin{equation}
\rho({\bf r}+{\bf a}_{i})=\rho({\bf r}) \, ,
\label{periodic}
\end{equation}
where ${\bf a}_{i}$, $i=1$, $2$, $3$, are a set of vectors which define a Bravais lattice. Therefore, 
$\rho({\bf r})$ can be expanded in a Fourier series:
\begin{equation}
\rho({\bf r})=\frac{1}{v}\sum_{\bf m}e^{-i{\bf k_{m}}\cdot{\bf x}}\, 
\hat{\rho}_{\bf m} 
\, ,
\label{fsum}
\end{equation}
where $v$ is the volume of the unit cell, ${\bf k_{m}}$ is a vector of the reciprocal lattice, 
and ${\bf m}$ denotes a set of three indexes $m_{i}=0,\pm 1,\pm 2\ldots$.
The expansion coefficients $\hat{\rho}_{\bf m}$ are given by:
\begin{equation}
\hat{\rho}_{\bf m}=\int_{v}\!\!d^{3}{\bf r}\, e^{i{\bf k_{m}}\cdot{\bf r}}
\rho({\bf r}) \, . 
\label{ftras}
\end{equation}
By use of Eqs.~(\ref{periodic}) and (\ref{fsum}), the functional~(\ref{dft}) can be rewritten 
in the following form:
\begin{equation}
\frac{\beta\Omega}{V}=
\frac{1}{v}\int_{v}\!\!d^{3}{\bf r}\, \rho({\bf r})
\left\{\ln[\rho({\bf r})]-1-\beta\mu\right\}+
\frac{1}{2 v^{2}}\sum_{\bf m}
|\hat{\rho}_{\bf m}|^{2} \, 
\beta\tilde{v}(\bf k_{m}) \, .
\label{fnew}
\end{equation}
We observe that, both in Eq.~(\ref{ftras}) and~(\ref{fnew}), the integration in ${\bf r}$ is restricted
to the unit cell. Hence, we may set ${\bf r}={\bf A}\cdot{\bf s}$, where ${\bf A}\equiv 
({\bf a}_{1},{\bf a}_{2},{\bf a}_{3})$ is the matrix obtained by arranging the lattice vectors ${\bf a}_{i}$ 
into columns, and ${\bf s}$ is a vector whose components vary in the interval $[-1/2, 1/2)$.   
By doing so, it is readily seen that in Eq.~(\ref{fnew}) neither the ideal-gas term nor the
Fourier components of the density profile $\hat{\rho}_{\bf m}$ that appear in the excess term depend 
on the specific kind of lattice: that is, these quantities are determined solely by the values 
$\rho({\bf s})\!\equiv\!\rho\, (\!{\bf A}\cdot{\bf s})$ assumed by the density profile in the unit cell, 
irrespective of the cell geometry. The information on the lattice enters in Eq.~(\ref{fnew}) {\em only}
via the reciprocal lattice vectors ${\bf k_{m}}$ at which $\tilde{v}(k)$ is evaluated.     
This feature makes it easy to implement a numerical procedure, in which the optimization of the grand
potential functional~(\ref{fnew}) is performed with respect to both $\rho({\bf s})$ and the cell geometry,
i.e., the elements of ${\bf A}$. In this study, we have assumed 
that the vectors ${\bf a}_{i}$ of the unit cell are mutually orthogonal, so that ${\bf A}$ is diagonal
with eigenvalues $2\pi/h_{i}$, and the reciprocal lattice vectors ${\bf k_{m}}$ have the form 
${\bf k_{m}}=(h_{1}m_{1},h_{2}m_{2},h_{3}m_{3})$. This assumption simplifies the calculation, but it could
be released without introducing any conceptually new element.     
To perform the minimization, one has to solve the Euler-Lagrange equations 
$\delta(\beta\Omega/V)/\delta\rho({\bf r})=0$ as well as $\partial(\beta\Omega/V)/\partial h_{i}=0$, 
where the functional derivative with respect to $\rho({\bf r})$ and the partial derivative with respect
to $h_{i}$ are given by:  
\begin{eqnarray}
& & \frac{\delta}{\delta\rho({\bf r})}\!\left(\frac{\beta\Omega}{V}\right) = 
\frac{1}{v}\left\{\ln[\rho({\bf r})]-\beta\mu\right\}+\frac{1}{v^{2}}
\sum_{\bf m}e^{-i{\bf k_{m}}\cdot{\bf r}}\hat{\rho}_{\bf m}\, 
\beta\tilde{v}({\bf k_{m}}) \, ,
\label{app:domegadrho}   \\
& & \frac{\partial}{\partial h_{i}}\!\left(\frac{\beta\Omega}{V}\right) = \frac{1}{v^{2}}
\sum_{\bf m}|\hat{\rho}_{\bf m}|^{2}
\frac{d\beta\tilde{v}}{d(k^{2})}({\bf k_{m}})\, h_{i}m_{i}^{2} \, .
\label{app:domegadh}
\end{eqnarray}
In the numerical solution, the functional $\Omega[\rho({\bf r})]$ was first discretized by sampling 
$\rho({\bf r})$ on a finite set of points $\rho_{\bf n}$, so as the replace the functional derivatives
with the partial derivatives with respect to $\rho_{\bf n}$. 
The minimization was then carried out by an iterative algorithm based on the steepest descent. 
In the basic version of the steepest descent, $\rho_{\bf n}$ and $h_{i}$ are updated recursively by moving 
``downhill'' 
in the direction opposite to that of the gradient of the discretized functional $\Omega_{\rm D}$:
\begin{eqnarray}
& & \rho_{\bf n}^{k+1} = \rho_{\bf n}^{k}-\eta\frac{\partial}{\partial\rho_{\bf n}}\!\left.\left(
\frac{\beta\Omega_{\rm D}}{V}\right)\right|_{k} \, ,  
\label{steep1} \\
& & h_{i}^{k+1} = h_{i}^{k}-\theta\frac{\partial}{\partial h_{i}}\!\left.\left(
\frac{\beta\Omega_{\rm D}}{V}\right)\right|_{k} \, ,
\label{steep2}  
\end{eqnarray}
where $k$ is the iteration index, and $\eta$, $\theta$ are the parameters which determine the size of
the downhill step. In order to increase its efficiency, the above algorithm was improved
by introducing preconditioning and conjugate gradients in Eq.~(\ref{steep1}), and by determining 
the step-size parameters $\eta$, $\theta$ adaptively at each iteration. A detailed description of these 
technical features will be given elsewhere~\cite{gaussmix}.         

The discretization of the density profile $\rho({\bf r})$ inside the unit cell was performed on 
$2^{7}\times 2^{7}\times 2^{7}=2\ 097\ 152$ points. The cell in real space was initially chosen as a cube 
with edge lengh $2\pi/h_{i}=10\sigma$, and was then evolved according to Eq.~(\ref{steep2}). 
The trial density profile $\rho_{\rm trial}({\bf r})$ used to start the minimization 
at a given thermodynamic state was set either to a random noise superimposed to a uniform
density, or to a sinusoidal modulation, or to the equilibrium $\rho({\bf r})$ of a nearby state. 
The need of performing different minimization runs for the same state is due to the fact that 
soft-core potentials such as that considered here may present many different inhomogeneous structures, 
each of which corresponds to a {\em local} minimum of the grand potential.    
Hence, the convergence of the algorithm does not guarantee {\em per se} that the {\em global} minimum 
has been identified: the search for the latter necessarily involves a comparison between 
different local minima of the grand potential. Of course, this problem is intrinsic to systems with 
a complex energy landscape, and many sophisticated strategies have been developed to tackle it~\cite{genetic,pickard}.
However, at the present stage we have not tried to implement any of them, 
and have just been content with our rather naive approach of performing different runs by changing 
$\rho_{\rm trial}({\bf r})$.         

\section{Results}
\label{sec:results}

\begin{figure}
\includegraphics[width=9cm]{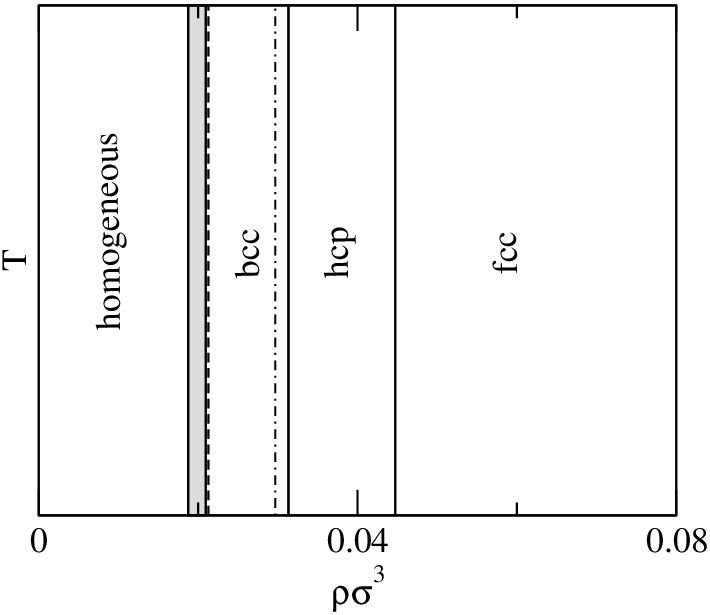}
\caption{Phase diagram of the potential $v_{\rm D}(r)$ of Eq.~(\ref{dendr}) obtained by the grand potential 
functional~(\ref{dft}). 
The gray stripe is the coexistence region between the homogeneous fluid and the bcc phase. The dash-dotted
line is the $\lambda$-line (see Sec.~\ref{sec:theory}). The dashed line gives the density of the fluid-solid 
transition according to Eq.~(\ref{rholikos}). The solid lines correspond to the transition between the bcc and hcp
phases and to that between the hcp and fcc phases. For both transitions, the coexistence region is too narrow 
to be resolved on the scale of the figure.}  
\label{fig:coex}
\end{figure}

By minimizing the grand potential~(\ref{dft}) via the algorithm described above it is found that, as expected, 
the system becomes inhomogeneous at sufficiently high density $\rho$. The phase diagram which we have obtained
is shown in Figure~\ref{fig:coex}. Since the potential is athermal, the temperature does not play any role 
in the phase behavior, which is then determined by the density alone. 
As one moves away from the homogeneous
region by increasing $\rho$, the fluid freezes by forming a cluster bcc phase. As anticipated in 
Sec.~\ref{sec:theory}, the $\lambda$-line actually lies inside the inhomogeneous region. The freezing and melting lines have been 
determined by performing a Maxwell construction, and the corresponding coexistence region is a rather narrow stripe 
centered at $\rho\sigma^{3}=0.020$. A calculation based on the same free-energy functional used 
here gives for the density of the fluid-solid transition of $Q^{\pm}$ potentials the expression~\cite{likoschem}:  
\begin{equation}
\rho=[\, 1.393\beta|\tilde{v}(k_{\rm m})|\, ]^{-1} \, . 
\label{rholikos}
\end{equation}
By substituting the above-mentioned value 
$\beta\tilde{v}_{\rm D}(k_{\rm m})=-33.686$, this gives $\rho\sigma^{3}=0.021$, in close agreement with 
the present result.

If $\rho$ is further increased, two more transitions are found, first from the bcc to a hcp phase, then from 
the hcp to a fcc phase. In both cases the densities of the coexisting phases are very close to each other, so that 
the width of the coexistence region cannot be resolved on the scale of Figure~\ref{fig:coex}.  
The transition from a bcc lattice to a close-packed lattice on increasing $\rho$ is in agreement with the general 
argument put forth in~\cite{likoschem}, as well as with the phase diagram of the GEM potential~(\ref{gem}) obtained 
in~\cite{likosrev}, although in that case the hcp phase was not observed.   

\begin{figure}
\includegraphics[width=9cm]{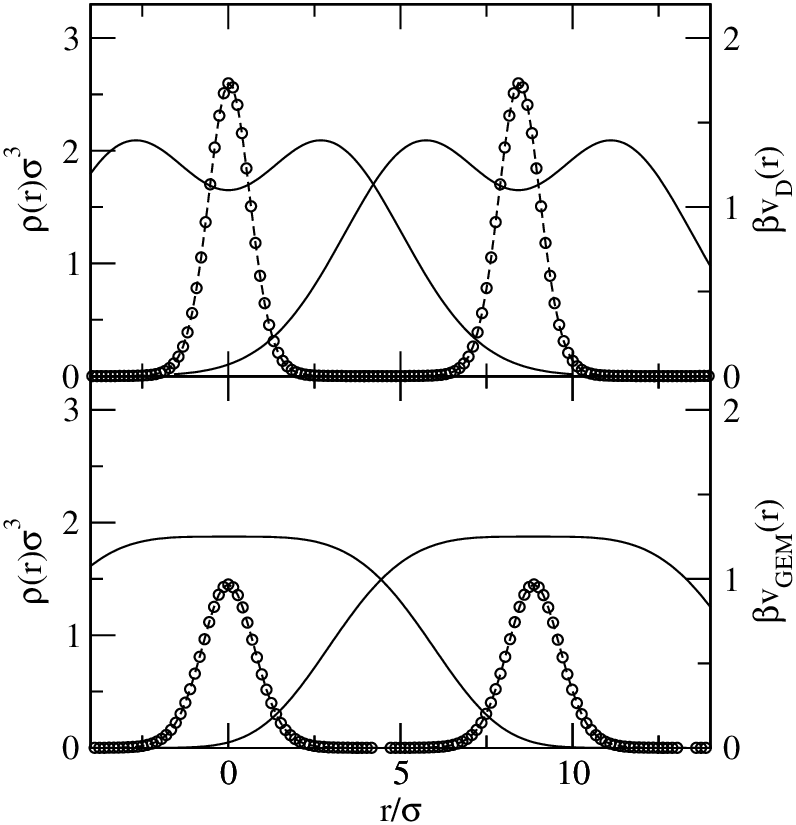}
\caption{Open circles: density profile $\rho({\bf r})$ along the direction connecting nearest-neighbor sites for 
$v_{\rm D}(r)$ at $\rho\sigma^{3}=0.023$ (upper panel) and $v_{\rm GEM}(r)$ at $\rho\sigma^{3}=0.022$ (lower panel)
with potential parameters determined as explained in the text. The dashed lines are a guide for the eye. 
Solid lines: dimensionless potentials $\beta v_{\rm D}(r)$ (upper panel) and $\beta v_{\rm GEM}(r)$ (lower panel) 
at each lattice site. The lattice is bcc for both potentials.}    
\label{fig:peakbcc}
\end{figure}

\begin{figure}
\includegraphics[width=9cm]{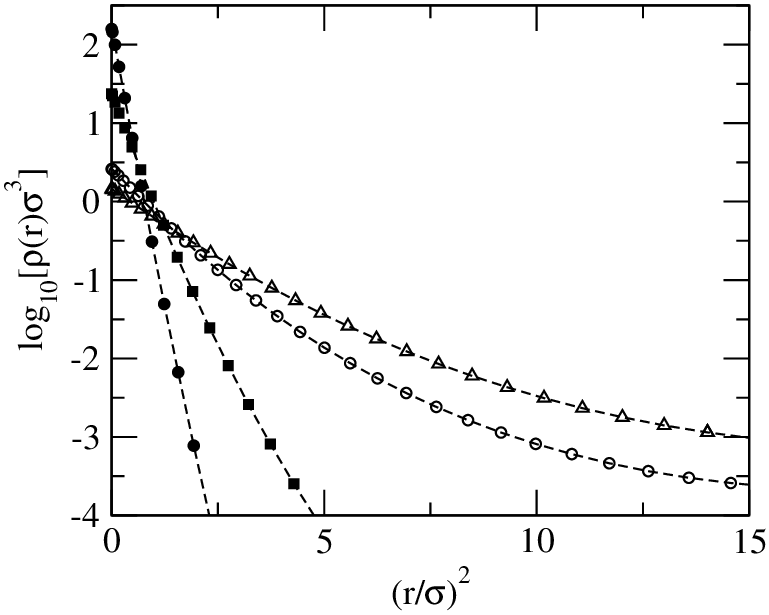}
\caption{Open symbols: logarithm of the density profiles displayed in the two panels of Figure~\ref{fig:peakbcc} 
for $v_{\rm D}(r)$ (open circles) and $v_{\rm GEM}(r)$ (triangles), as a function
of the square of the distance from a given lattice site. 
Filled symbols: same as the above for $v_{\rm D}(r)$ and $\rho\sigma^{3}=0.046$ (squares), 
$\rho\sigma^{3}=0.093$ (circles). The dashed lines are a guide for the eye.} 
\label{fig:peaklog}
\end{figure}

The density profile $\rho({\bf r})$ at a density $\rho\sigma^{3}=0.023$ in the bcc region near the fluid-solid 
boundary is shown 
in the upper panel of Figure~\ref{fig:peakbcc} along the direction connecting nearest neighbors as a function
of the distance $r$ from a given lattice site. 
In order to 
assess the spatial extension of the density peaks against that of the interaction potential, the latter has also been 
plotted in the figure. We observe that, despite the proximity to the fluid region, $\rho({\bf r})$ is already 
strongly localized at the lattice sites. It is interesting to compare this density profile with
that of the GEM potential at a similar density. To this purpose, we have considered the interaction~(\ref{gem}) 
at a reduced temperature
$k_{\rm B}T/\epsilon=0.8$ such that $\beta v_{\rm GEM}(r=0)=1.25$, somewhere between the well and the hump 
of $\beta v_{\rm D}(r)$, see Figure~\ref{fig:dendr}. The width $R$ of $v_{\rm GEM}(r)$ has then been fixed at the value
$R=6.4264\sigma$, for which the integrated intensity of $\beta v_{\rm GEM}(r)$ is the same as that 
of $\beta v_{\rm D}(r)$.
The lower panel of Figure~\ref{fig:peakbcc} shows $\rho({\bf r})$ obtained by the numerical minimization of 
the grand potential functional~(\ref{dft}) for the GEM potential at $\rho\sigma^{3}=0.022$, as well as 
$\beta v_{\rm GEM}(r)$ for the parameters given above. The most stable lattice is again the bcc, but 
the peaks are significantly lower. Since the density is about the same in both panels, this must be compensated
by a larger width of the peaks. That this is indeed the case is shown more clearly in Figure~\ref{fig:peaklog}, 
where the logarithm of the two density profiles of Figure~\ref{fig:peakbcc} has been plotted as a function of $r^{2}$. 
Hence, the dendrimer potential $v_{\rm D}(r)$ leads to a higher localization with respect to the GEM potential, 
and it is natural to trace this back to the minimum of $v_{\rm D}(r)$ at $r=0$.    
We also note that on the scale of Figure~\ref{fig:peaklog}, a Gaussian $\rho({\bf r})$ corresponds 
to a straight line. It appears that, for both $v_{\rm D}(r)$ and $v_{\rm GEM}(r)$, $\rho({\bf r})$ can be
accurately represented by a Gaussian as long as one has $\rho({\bf r})\sigma^{3}\lesssim 0.1$, as one would expect 
from Figure~\ref{fig:peakbcc}. Deviations from the Gaussian profile become evident at smaller $\rho({\bf r})$.     

\begin{figure}
\includegraphics[width=9cm]{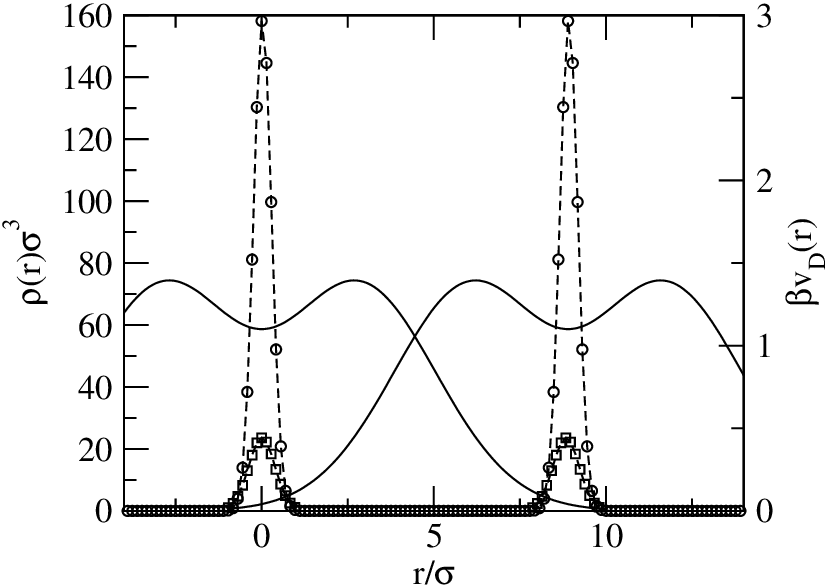}
\caption{Same as the upper panel of Figure~\ref{fig:peakbcc} for $\rho\sigma^{3}=0.046$ (squares) and 
$\rho\sigma^{3}=0.093$ (circles). The dashed lines are a guide for the eye. The lattice is fcc for both densities.}
\label{fig:peakfcc}
\end{figure}

As one moves more deeply into the inhomogeneous region, the height of the peaks of the density profile increases very 
rapidly, as shown in Figure~\ref{fig:peakfcc} which displays $\rho({\bf r})$ in the fcc domain for 
$\rho\sigma^{3}=0.046$ and $\rho\sigma^{3}=0.093$. These $\rho({\bf r})$ have also been plotted 
in Figure~\ref{fig:peaklog} in order to show that the peaks become narrower and narrower on increasing $\rho$.     
On the other hand, by comparing Figures~\ref{fig:peakbcc} and~\ref{fig:peakfcc} it appears that the nearest-neighbor 
distance
$d$ is hardly affected by density. In fact, this behavior is peculiar of crystals formed by $Q^{\pm}$ potentials 
for which, even though the specific kind of periodicity of $\rho({\bf r})$  
depends of course on the lattice formed by the system, its spatial extent is basically determined 
only by the density-independent wave vector $k_{\rm m}$ of the minimum of the potential in Fourier 
space~\cite{likoschem}. At variance
with atomic crystals, this leads to a nearest-neighbor distance $d$ and a volume per lattice site $v_{0}$  
which are nearly unaffected by $\rho$. Since the number of particles 
$n_{c}$ in a cluster located at a lattice site is related to the average density $\rho$
by the relation~\cite{likoschem} $\rho=n_{c}/v_{0}$, a nearly constant $v_{0}$ implies 
a nearly linear dependence of $n_{c}$ on $\rho$.  

\begin{figure}
\includegraphics[width=9cm]{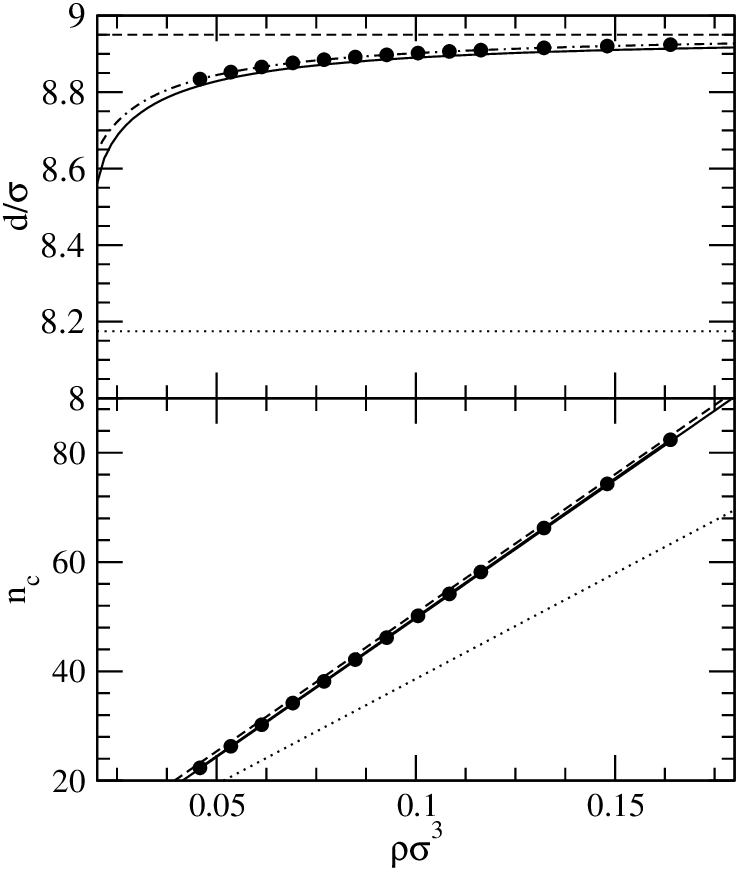}
\caption{Nearest-neighbor distance $d$ (upper panel) and number of particles in a cluster $n_{c}$ (lower panel) for the
potential $v_{\rm D}(r)$ in the fcc phase. The filled circles are the result obtained by the numerical minimization 
of functional~(\ref{dft}). The dotted lines are the result obtained by estimating $d$ from the wave vector 
$k_{\rm m}$ of the minimum of $\tilde{v}_{\rm D}(k)$. The dashed and solid lines have been obtained by 
representing the density profile by a superposition 
of $\delta$-like peaks or Gaussians respectively, and taking into account only onsite and nearest-neighbor 
interactions. The dot-dashed line in the upper panel has been obtained by the Gaussian representation of the
density profile and inclusion of the interactions beyond nearest neighbors, see Sec.~\ref{sec:gaussian}.}  
\label{fig:cluster}
\end{figure}

The nearest-neighbor distance and the number of particles per cluster are shown in Figure~\ref{fig:cluster} for the fcc 
phase. Although not strictly constant, $d$ is indeed found to be a very slowly varying function of density, giving 
the anticipated linear dependence of $n_{c}$ on $\rho$. It is interesting to observe that, as $\rho$ is increased,     
$d$ is actually found to increase, albeit slightly so. At first this may appear at odds with the fact that, if the 
particle number $N$ is kept constant and $\rho$ is increased, the volume of the system must obviously decrease, 
but in fact no contradiction arises, since the slight increase in $v_{0}$ is more than compensated by the decrease
in the number of lattice sites $N_{l}=N/n_{c}$. In other words, a cluster crystal manages
to change its volume not by changing the lattice spacing, but rather the number of lattice sites, and consequently 
their population $n_{c}$. In this respect, whether $d$ increases or decreases on increasing $\rho$ is of little 
relevance, provided it remains a slowly varying function of $\rho$.    

\section{Analytical representation of $\rho({\bf r})$}
\label{sec:gaussian}

As evident from Figures~\ref{fig:peakbcc} and~\ref{fig:peakfcc}, as $\rho$ increases the density profile becomes
very strongly localized at the lattice sites. It is then natural to represent it by a sum of non-overlapping
or weakly overlapping peaks centered at these sites. The crudest approximation of this kind consists in 
assuming the highest degree of localization, and modeling $\rho({\bf r})$ as a sum of $\delta$-functions:
\begin{equation}
\rho_{\delta}({\bf r})=n_{c}\sum_{\bf m}\delta({\bf r}-{\bf R_{m}}) \, ,
\label{rhodelta}
\end{equation}  
where ${\bf R_{m}}$ describe the positions of the lattice sites. Strictly speaking, Eq.~(\ref{rhodelta}) cannot 
be true, since it would imply an infinite free energy cost due to the loss of translational entropy. 
In a system with temperature-independent interactions, Eq.~(\ref{rhodelta}) would hold at $T=0$, but then 
$n_{c}$ could not be chosen at one's will, since it would be bounded to assume integer values~\cite{likosj}.  
However, the above assumption 
still makes sense as an approximation, if one assumes that, for the true $\rho({\bf r})$, the free energy is 
dominated by its energetic contribution and may thus be approximated by disregarding the entropic term altogether.  
One is then left with an internal energy per particle $(E/N)_{\delta}$ which, for $\rho({\bf r})$ 
of Eq.~(\ref{rhodelta}), has the form:
\begin{equation}
\left(\frac{\beta E}{N}\right)_{\!\delta}=\frac{1}{2}\zeta d^{3}\!\rho [\, \beta v_{\rm D}(0)
+\sum_{m\neq 0}q_{m}\beta v_{\rm D}(c_{m}d)\, ]-\frac{1}{2}\beta v_{\rm D}(0) \, . 
\label{madelung}
\end{equation}
In the above expression, $v_{\rm D}(0)$ represents the interaction between particles belonging to the same cluster,
and the interaction between different clusters is given by the lattice sum of the potential over 
successive neighbor shells. The constants $q_{m}$ and $c_{m}$ are specific to the lattice considered, and represent
respectively the number of $m$-th neighbors and their distance from a given lattice site in units 
of the nearest-neighbor distance $d$. The number of particles per cluster $n_{c}$ has been expressed as 
$\zeta d^{3}\rho$, where the quantity $\zeta d^{3}$ is the volume per lattice site $v_{0}$ 
expressed as a function of $d$, and again $\zeta$ depends on the lattice: in particular, $\zeta\!=\!4\sqrt{3}/9$ 
for the bcc lattice and $\zeta\!=\!\sqrt{2}/2$ for the fcc and hcp lattices.
If Eq.~(\ref{madelung}) is truncated at the nearest-neighbor level, one obtains:
\begin{equation}
\left(\frac{\beta E}{N}\right)_{\!\delta}\simeq \frac{1}{2}\zeta d^{3}\!\rho
\left[\, \beta v_{\rm D}(0)+q \beta v_{\rm D}(d)\, \right] \, ,
\label{madelungapp}
\end{equation}
where $q\equiv q_{1}$ is the number of nearest neighbors. Equation~(\ref{madelungapp}) can be regarded as 
a simple variational expression to be minimized at fixed $\rho$ with respect to the single variational parameter 
$d$. Again, we must observe that this statement taken at face value makes little sense, since the minimum 
of Eq.~(\ref{madelungapp}) at fixed $\rho$ is actually obtained for $d=0$: in other words, this expression does
not prevent the unphysical collapse of the crystal at a single lattice point. What drives the system away 
from such a collapse is the fact that, as $d$ is decreased, more and more shells 
contribute significantly to the energy. This leads to a divergence of the sum in Eq.~(\ref{madelung}), but 
such an occurrence is lost in Eq.~(\ref{madelungapp}), which contains just one shell. Nevertheless, 
Figures~\ref{fig:peakbcc} and~\ref{fig:peakfcc} indicate that, at the {\em actual} $d$, the interaction energy 
between different lattice sites is rapidly decreasing with their distance, so that the contribution
due to the outer shells should indeed be very small. Hence, although Eq.~(\ref{madelungapp}) cannot be trusted  
at every $d$, it might still be a reasonable approximation in the neighborhood of the value of $d$ chosen by
the system, which would then be recovered as a {\em local} minimum of Eq.~(\ref{madelungapp}). 
In fact, besides the spurious minimum at $d=0$, this expression is found to have a (spurious) local maximum and 
a local minimum. The latter is equal to $d_{\rm bcc}=8.35\sigma$ for the bcc lattice 
and $d_{\rm fcc}=8.95\sigma$ for the fcc and hcp lattices. In Figure~\ref{fig:cluster} we have reported $d_{\rm fcc}$
together with the value $d_{\lambda}=\pi\sqrt{6}/k_{\rm m}$ obtained by identifying the wave vector $k_{\rm m}$ 
of the minimum of $\tilde{v}_{\rm D}(k)$ with the nearest-neighbor distance in reciprocal space, 
which yields $d_{\lambda}=8.17$.
One sees that $d_{\rm fcc}$ is a rather accurate estimate of the nearest-neighbor distance, actually more so than 
$d_{\lambda}$. As a consequence, also the slope of the $n_{c}$ vs. $\rho$ plot is accurately reproduced.   

The value of $d$ obtained from the minimization of $(E/N)_{\delta}$ is clearly independent of density,
and this would be true even if the summation over all neighbor shells of Eq.~(\ref{madelung}) were used instead of
the truncated expression of Eq.~(\ref{madelungapp}). In order to describe the dependence of $d$ on density, 
the contribution of the translational entropy to the free energy must be taken into account, and this in turn 
requires to go beyond the representation of $\rho({\bf r})$ as a superposition of $\delta$-like spikes.  
Following Refs.~\cite{likosrev} and~\cite{likoschem}, we then replace the $\delta$-functions in Eq.~(\ref{madelung})
with Gaussians of finite amplitude:
\begin{equation}
\rho_{G}({\bf r})=n_{c}\left(\frac{\alpha}{\pi}\right)^{3/2}\sum_{\bf m}e^{-\alpha({\bf r}-{\bf R_{m}})^{2}} \, ,
\label{rhogauss}
\end{equation}
where the normalization constant has been determined so that $n_{c}$ gives the number of particles at each site 
as requested. If $\rho_{G}({\bf r})$ is substituted into Eq.~(\ref{dft}) and the overlap between the Gaussians
at different lattice sites is neglected in the ideal-gas term, then the following expression for 
the Helmholtz free energy $F\equiv\Omega+\mu\int\!d^{3}{\bf r}\rho({\bf r})$ is obtained~\cite{likosrev}:   
\begin{eqnarray}
& & \left(\frac{\beta F}{N}\right)_{\!\rm G}\!=\ln n_{c}+\frac{3}{2}\ln\frac{\alpha}{\pi}-\frac{5}{2}
+n_{c}\sqrt{\frac{\alpha^{3}}{2\pi}}\int_{0}^{+\infty}\!\!\!dr\, r^{2}e^{-\alpha r^{2}/2}\, \beta v(r) 
\nonumber \\
& & \mbox{}+ n_{c}\sqrt{\frac{\alpha}{8\pi}}\sum_{{\bf m}\neq 0}\frac{1}{R_{\bf m}}\int_{0}^{+\infty}\!\!\!dr\, r
\left[e^{-\alpha(r-R_{\bf m})^{2}/2}-e^{-\alpha(r+R_{\bf m})^{2}/2}\right]\beta v(r) \, .
\label{freegauss}
\end{eqnarray}
For the potential $v_{\rm D}(r)$, the integrals in Eq.~(\ref{freegauss}) can be performed analytically, 
and one has:
\begin{eqnarray} 
& &\mbox{\hspace{-1.2cm}}
\left(\frac{\beta F}{N}\right)_{\!\rm G}\!= \ln(\zeta d^{3}\rho) +\frac{3}{2}\ln\frac{\alpha}{\pi}-\frac{5}{2} 
+\frac{1}{2}\, \zeta d^{3}\rho\left\{\epsilon_{1}\!\left(\frac{\alpha}{\alpha+\gamma}\right)^{3/2}\!
\{1+\sum_{m\neq 0}q_{m}e^{-\alpha\gamma c_{m}^{2}d^{2}/[2(\alpha+\gamma)]}\, \}\right.
\nonumber \\
& & \left.- \epsilon_{2}\!\left(\frac{\alpha}{\alpha+\delta}\right)^{3/2}\!
\{1+\sum_{m\neq 0}q_{m}e^{-\alpha\delta c_{m}^{2}d^{2}/[2(\alpha+\delta)]}\, \}\right\}  \, , 
\label{freedendr}
\end{eqnarray}
where $\gamma=2/R_{1}^{2}$, $\delta=2/R_{2}^{2}$, and we have expressed $n_{c}$ as $n_{c}=\zeta d^{3}\rho$ 
as before. In the limit $\alpha\rightarrow\infty$, the interaction 
term in Eq.~(\ref{freedendr}) reproduces Eq.~(\ref{madelung}), save for the thermodynamically irrelevant term  
$-\beta v_{\rm D}(0)/2$. However, the minimization of the free energy $(\beta F/N)_{\rm G}$ must now be carried out 
with respect to both $d$ and $\alpha$. This requires the solution of the non-linear system of equations
$\partial(\beta F/N)_{\rm G}/\partial d=\partial(\beta F/N)_{\rm G}/\partial\alpha=0$. If we make the change  
of variables: 
\begin{eqnarray}
x & = & \frac{\gamma d^{2}}{2(1+\gamma/\alpha)} \, , \nonumber \\
y & = & \frac{\delta d^{2}}{2(1+\delta/\alpha)} \, , 
\label{change}
\end{eqnarray}
the above system can be cast in the rather symmetric form:
\begin{eqnarray}
x & = & H \frac{\gamma}{\delta}\, \frac{y S(y)}{H S(y)-3} \, , \nonumber \\
y & = & K \frac{\delta}{\gamma}\, \frac{x S(x)}{K S(x)+3} \, ,
\label{system}
\end{eqnarray}
where we have set: 
\begin{eqnarray}
H & = & \sqrt{2}\, \zeta\rho\frac{\epsilon_{2}}{\delta^{3/2}} \, ,  \nonumber \\
K & = & \sqrt{2}\, \zeta\rho\frac{\epsilon_{1}}{\gamma^{3/2}} \, ,
\label{coeff}
\end{eqnarray}
and the function $S(x)$ is defined by:
\begin{equation}
S(x)=x^{3/2}\left[3+\sum_{m\neq 0}q_{m}(3-2c_{m}^{2}x)e^{-c_{m}^{2}x}\right] \, .
\label{sum}
\end{equation}
We can now introduce the same approximation used in going from Eq.~(\ref{madelung}) to Eq.~(\ref{madelungapp}) 
by truncating $S(x)$ at the nearest-neighbor shell:
\begin{equation}
S(x)\simeq x^{3/2}\left[3+q(3-2x)e^{-x}\right] \, .
\label{sumapp}
\end{equation}
As observed above for Eq.~(\ref{madelungapp}), strictly speaking such a truncation would lead to the collapse
of the crystal at $d=0$, but one expects that a physical solution may still be recovered as a local minimum of
the free energy thus obtained. The results for $d$ and $n_{c}$ given by the numerical solution of Eq.~(\ref{system})
with the approximation~(\ref{sumapp}) for the fcc lattice are again displayed in Figure~\ref{fig:cluster}. 
Both $d$ and $n_{c}$ run very close to the corresponding quantities obtained from the fully numerical
minimization of the free energy described in Sec.~\ref{sec:theory}. It appears then that, provided the finite amplitude 
of the density peaks is taken into account,
keeping the interactions only up to nearest neighbors is already sufficient to reproduce accurately the density 
dependence of $d$, at least at high density where the fcc phase is favored. For the bcc phase, it turns out that also the second-nearest neighbor shell must be included in the calculation. This is due to the fact that in the bcc lattice the nearest and second-nearest neighbor shells are much closer to each other than in the fcc lattice. In either the fcc or the bcc, inclusion of the other neighbor shells from the third on leads to a result which is indistinguishable from that given
by the fully numerical calculation, as again shown in the figure for the fcc phase.      

\begin{figure}
\includegraphics[width=9cm]{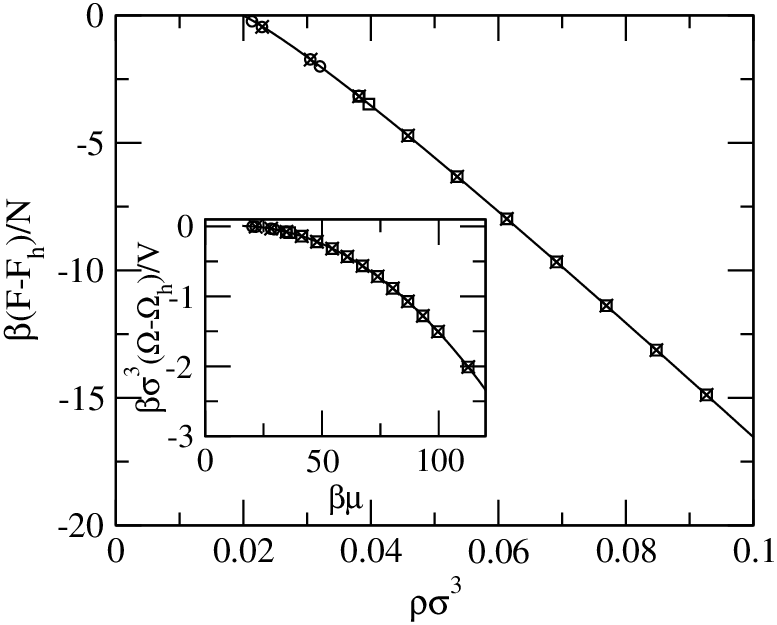}
\caption{Helmholtz free energy per particle $F/N$ as a function of density in the inhomogeneous region, measured 
with respect to that of the homogeneous fluid $F_{\rm h}/N$. The symbols refer to the results of the fully numerical
minimization for the bcc (circles), fcc (squares), and hcp phases (crosses). The solid line is
the result obtained by representing $\rho({\bf r})$ by a superposition of Gaussians, see Eqs.~(\ref{rhogauss}) and 
(\ref{freedendr}). In both cases, the free energy differences between different phases cannot be resolved on the scale 
of the figure. 
Inset: same as the above for the grand potential per unit volume $\Omega/V$ measured with respect to that 
of the homogeneous fluid $\Omega_{\rm h}/V$.}    
\label{fig:free}
\end{figure}

In view of such an agreement, it is natural to ask how the phase behavior predicted by Eq.~(\ref{freedendr}) 
on the basis of the Gaussian parameterization of the density profile~(\ref{rhogauss}) compares with that discussed 
at the beginning of Sec.~\ref{sec:results}. This is a rather subtle issue, because the free energies 
of the different phases are very close to one another. Therefore, even a small change in their values may alter their 
relative balance. 
In fact, it is found that the truncation of the interactions at the nearest-neighbor level in Eq.~(\ref{freedendr})
has rather important consequences on the phase diagram, because it would make the bcc phase more stable than both 
the fcc and hcp phases even at high density, thus leaving the bcc as the only inhomogeneous phase. In order 
to recover the stability of the close-packed structures at high density, in Eq.~(\ref{freedendr}) the interactions have
to be taken into account at least up to the second-nearest neighbor shell. As recalled above, the contribution of this shell is much more substantial in the bcc lattice than in either the fcc or hcp, and proves crucial in order to move the bcc free energy above that of the close-packed lattices.  
In Figure~\ref{fig:free} we show 
the free energy per particle $F/N$ and the grand potential per unit volume $\Omega/V$ of the bcc, fcc and hcp phases 
obtained via the minimization of Eq.~(\ref{freedendr}) 
by including a rather large number of neighbor shells such that, for the densities considered, the addition 
of further shells would leave the free energy unchanged within numerical accuracy. 
These results are compared to those 
given for the same phases by the fully numerical minimization. It is evident that the two sets of data are 
indistinguishable on the scale of the figure. Moreover, for both sets of data the free energies and grand potentials 
corresponding to different phases are also indistinguishable, although by enlarging the energy scale the small 
differences which determine the most stable phase at a certain density would be appreciated. In particular, 
Eq.~(\ref{freedendr}) predicts that the bcc lattice becomes unfavorite with respect to close-packed lattices 
for $\rho\sigma^{3}>0.031$, in very good agreement with the position of the bcc-hcp transition shown 
in Figure~\ref{fig:coex}. 

There is, however, a difference between the phase behavior 
given by the unconstrained minimization discussed in Sec.~\ref{sec:results} and that given by Eq.~(\ref{freedendr}), 
namely, while the former predicts the occurrence of the hcp phase between the bcc and fcc phases, according 
to the latter the free energy of the hcp phase is always higher than that of the fcc phase. Hence, the hcp phase is
not observed, and as the density is increased the sequence of phase transitions is fluid--bcc--fcc instead of
fluid--bcc--hcp--fcc as in Figure~\ref{fig:coex}. Nevertheless, it should be remarked that in both cases the free 
energies of the hcp and fcc phases are everywhere extremely close to each other: specifically, their relative
difference is of the order of $\sim 10^{-6}$, while that between the bcc and fcc phases is $\sim 10^{-3}$. 
This is not surprising, in the light of the fact that the free energies of the fcc and hcp 
lattices become distinct only from the third-neighbor shell on. Therefore, even though the numerical accuracy of the
calculation is high enough to allow resolution of such tiny differences, speaking of a ``most stable'' phase    
in this context is somewhat academic, and one should more realistically regard the fcc and hcp phases as nearly
degenerate. One also expects that whether the contributions to the free energy from the high-order shells 
will stabilize one phase or the other will most likely depend on rather specific features of the decay of the potential 
at long distance.  

\section{Conclusions}
\label{sec:concl}

We have employed density-functional theory to study the phase diagram and the density profile of a system of particles 
interacting via an athermal, penetrable-core potential $v_{\rm D}(r)$ aimed at modeling the effective pair interaction 
between dendrimers~\cite{lenz}. The potential belongs to the so-called $Q^{\pm}$ class~\cite{likosq}, which is expected 
to lead 
to the formation of cluster crystals at high enough density. The results discussed in the paper confirm this prediction,
and the main features of the behavior of the system are those already encountered in the 
study of the generalized exponential model potential $v_{\rm GEM}(r)=\epsilon\exp[-(r/R)^{4}]$ carried out some years 
ago~\cite{likoschem,likosrev}. Specifically: i) in the inhomogeneous phases, the particles arrange into clusters localized
at the lattice sites, such that each cluster contains many particles. ii) The lattice spacing is nearly 
independent of density, which in turn implies that the number of particles per cluster is a nearly linear function 
of density. In the present system the localization is even stronger than that observed in the GEM potential, 
because of the local minimum presented by $v_{\rm D}(r)$ at $r=0$. iii) As the density is increased, one observes 
a transition from the bcc structure to close-packed structures. Here we find both hcp and fcc phases, 
while for the GEM potential the hcp phase was not observed. 

The density functional which we have adopted has the same mean-field form already used in the study 
of the GEM potential~\cite{likoschem,likosrev}. The minimization of the functional has been performed 
by a parameter-free procedure, in which both the values assumed by the density profile $\rho({\bf r})$ 
and the lattice structure were determined as the outcome of the calculation. The only assumptions which we made are 
that the density profile is periodic, and that the {\em unit} cell can be represented 
by a set of mutually orthogonal vectors. Although not absolutely general, this assumption does not necessarily rule out 
the occurrence of structures whose {\em primitive} cell cannot be represented in this way, e.g. the aforementioned
hcp lattice. The results thus obtained have then been compared with those given by the more common procedure, 
in which one considers {\em a priori} a set of possible lattice structures, represents $\rho({\bf r})$ as 
a superposition of Gaussians centered at the lattice sites, and identifies  
the most stable structure among the selected ones by inspection of their free energies once the minimization has been 
carried out. The comparison shows that the Gaussian parameterization reproduces most of the results obtained 
by the unconstrained minimization. In fact, for the densities at which the system arranges into a close-packed lattice, its properties can be already described very satisfactorily by 
further simplifying the treatment, and disregarding the contribution to the excess free energy beyond nearest-neighbor
lattice sites, although this contribution is necessary in order to account for the transition from the bcc to the
fcc structure. The only qualitative prediction which we did not recover by the Gaussian 
representation of $\rho({\bf r})$ is the occurrence of the hcp phase between the bcc and fcc, the fcc phase being
always favored with respect to the hcp. However, in both approaches the relative differences in free energy between 
the two structures are extremely small, so that labelling either of them as the more stable one is of little practical
relevance. In fact, these differences are due to the interactions from the third nearest-neighbor shell on, which 
we expect to depend on rather specific details of the decay of the interaction at long distance. 
In the light of these considerations, for the system considered here the unconstrained minimization  
may admittedly appear as an overkill. One can easily imagine, however, that this may not be the case in other 
situations, in which the density profile is less amenable to being described by a superposition of Gaussians. 
We plan to report on such an instance in the near future~\cite{gaussmix}.    
       
Before concluding this discussion, one more consideration is in order:
since the potential $v_{\rm D}(r)$ has been introduced as a representation of pair interactions between dendrimers,
one may ask whether real dendrimers can be expected to display the behavior described above, especially as far as 
cluster formation is concerned. This depends on the actual relevance of many-body interactions, which are disregarded 
in the present picture. The very fact that, according to the density profile predicted on the basis of $v_{\rm D}(r)$,
most particles concentrate at the lattice sites leading to very high values of the local density, suggests 
that many-body effects are indeed relevant, and may significantly influence the behavior of the system. This conclusion
was reached in a computer simulation study of an assembly of ring polymers~\cite{narros}. For that system, the 
pair potential between two isolated polymers displays a profile similar to that of $v_{\rm D}(r)$. In particular,
it also belongs to the $Q^{\pm}$ class, so that it is expected to promote cluster formation. However, it was
found that at finite density many-body effects lead to an effective pair interaction which is substantially modified 
with respect to its zero-density limit, to the point that clustering does not take place. In such a situation, 
the pair interaction between two isolated ring polymers is clearly of little use in understanding the behavior 
of the system. On the other hand, a subsequent study showed that in dendrimers the formation of cluster crystals
does take place~\cite{lenz2}, although many-body effects are still very important. In fact, they prevent 
the number of dendrimers per lattice site $n_{c}$ from growing indefinitely, as is instead predicted by the two-body
potential, and inhibit the formation of the crystal at high density, where a percolated network of dendrimers 
is favored. Moreover, in line with this tendency of many-body interactions to curb the growth of $n_{c}$ 
with $\rho$, the lattice spacing was found to be a slowly decreasing function of $\rho$, rather than a slowly increasing
one as in the present study.
It should also be observed that the parameters of the interactions between monomers used 
in~\cite{lenz2} are different from those used in~\cite{lenz}, and the resulting pair potential 
for isolated dendrimers leads to clustering at a lower density than that obtained with the pair potential~(\ref{dendr}) 
considered here. Nevertheless, it is reassuring to have evidence that for this system 
the formation of cluster crystals does not rest solely upon the effective pair potential picture.  
 
\section*{Acknowledgements}    

This paper is dedicated to Johan H\o ye on his 70$^{\rm th}$ birthday.
The author wishes to thank Alberto Parola, Luciano Reatto, and Christos Likos for illuminating conversation. 




\begin{thebibliography}{99}

\bibitem{stillinger} F.~H.~Stillinger and D.~K.~Stillinger, {\it Physica A} {\bf 244}, 358 (1977);
A.~Lang, C.~N.~Likos, M.~Watzlawek and H.~L\" owen, {\it J. Phys.: Condens. Matter} {\bf 12}, 5087 (2000);
C.~N.~Likos, {\it Phys. Rep.} {\bf 348}, 267 (2001).    
%
\bibitem{narros} A.~Narros, A.~J.~Moreno, and C.~N.~Likos, {\it Soft Matter} {\bf 6}, 2435 (2010).
%
\bibitem{mladekrev2} B.~M.~Mladek, G.~Kahl, and C.~N.~Likos, {\it Phys. Rev. Lett.} {\bf 100}, 028301 (2008). 
%
\bibitem{lenz} D.~A.~Lenz, B.~M.~Mladek, C.~N.~Likos, G.~Kahl, and R.~Blaak, {\it J. Phys. Chem. B} {\bf 115},
7218 (2011).  
%
\bibitem{likosq} C.~N.~Likos, A.~Lang, M.~Watzlawek, and H.~L\"owen H {\it Phys. Rev. E} {\bf 63}, 031206 (2001).
%
\bibitem{likoschem} C.~N.~Likos, B.~M.~Mladek, D.~Gottwald, and G.~Kahl, {\it J. Chem. Phys.} {\bf 126}, 224502 (2007).
%
\bibitem{compe} R.~P.~Sear, S.-W.~Chung, G.~Markovich, W.~M.~Gelabrt, and J.~R.~Heath, {\it Phys. Rev. E} {\bf 59}, 
R6255 (1999); R.~P.~Sear and W.~M.~Gelbart, {\it J. Chem. Phys.} {\bf 110}, 4582 (1999);
A.~Imperio and L.~Reatto, {\it J. Phys.: Condens. Matter} {\bf 16}, S3769 (2004);
A.~Imperio and L.~Reatto, {\it J. Chem. Phys.} {\bf 124}, 164712 (2006);
A.~Imperio and L.~Reatto, {\it Phys. Rev. E} {\bf 76}, 040402 (2007);
A.~J.~Archer A J and N.~B.~Wilding, {\it Phys. Rev. E} {\bf 76}, 031501 (2007);
%
\bibitem{archer} A.~J.~Archer, {\it Phys. Rev. E} {\bf 78}, 031402 (2008).
%
\bibitem{likosrev} B.~M.~Mladek, D.~Gottwald, G.~Kahl, M.~Neumann, and C.~N.~Likos, {\it Phys. Rev. Lett} {\bf 96},
045701 (2006).
%
\bibitem{mladekrev} B.~M.~Mladek, P.~Charbonneau, and D.~Frenkel, {\it Phys. Rev. Lett.} {\bf 99}, 235702 (2007). 
%
\bibitem{likosj} T.~Neuhaus and C.~N.~Likos, {\it J. Phys.: Condens. Matter} {\bf 23}, 234112 (2011). 
%
\bibitem{zhang} K.~Zhang, P.~Charbonneau, and B.~M.~Mladek, {\it Phys. Rev. Lett.} {\bf 105}, 245701 (2010);
N.~B.~Wilding and P.~Sollich, {\it EPL} {\bf 101},10004 (2013).  
%
\bibitem{lenz2} D.~A.~Lenz, R.~Blaak, C.~N.~Likos, and B.~M.~Mladek, {\it Phys. Rev. Lett.} {\bf 109}, 228301 (2012).
%
\bibitem{mladekj} B.~M.~Mladek, D.~Gottwald, G.~Kahl, M.~Neumann, and C.~N.~Likos, {\it J. Chem. Phys. B} {\it 111},
12799 (2007).
%
\bibitem{gaussmix} D.~Pini, A.~Parola, and L.~Reatto, in preparation.
%
\bibitem{genetic} D.~Gottwald, G.~Kahl, and C.~N.~Likos, {\it J. Chem. Phys.} {\bf 122}, 204503 (2005).
%
\bibitem{pickard} C.~J.~Pickard and R.~J.~Needs, J. Phys.: {\it Condens. Matter} {\bf 23}, 053201 (2011).



\end{thebibliography}
\end{document}